\documentstyle[amsfonts,12pt]{article}

\title{Proof of the Julia--Zee Theorem}
\author{Joel Spruck$^{1}$ and Yisong Yang$^{2}$\\
$^1${\small{\em Department of Mathematics, Johns Hopkins University, Baltimore, Maryland 21218}}\\
$^2${\small{\em Department of Mathematics, Yeshiva University, New
York, New York 10033, USA}}
}
\date{}
\newcommand{\bfR}{{\Bbb R}}

\newcommand{\bA}{{{\bf A}}}\newcommand{\bF}{{\bf F}}

\newtheorem{theorem}{Theorem}[section]
\newtheorem{oldtheorem}{Theorem}[section]
\newtheorem{oldassertion}[oldtheorem]{Assertion}
\newtheorem{oldproposition}[oldtheorem]{Proposition}

\newtheorem{oldlemma}[oldtheorem]{Lemma}
\newtheorem{olddefinition}[oldtheorem]{Definition}
\newtheorem{oldclaim}[oldtheorem]{Claim}
\newtheorem{oldcorollary}[oldtheorem]{Corollary}

\newenvironment{proposition}{\begin{oldproposition}$\!\!\!${\bf.}}
{\end{oldproposition}}

\newbox\qedbox
\newenvironment{proof}{\smallskip\noindent{\bf Proof.}\hskip \labelsep}%
                        {\hfill\penalty10000\copy\qedbox\par\medskip}

\newcommand{\dd}{\mbox{d}}
\newcommand{\ee}{\end{equation}}
\newcommand{\be}{\begin{equation}}
\newcommand{\bea}{\begin{eqnarray}}
\newcommand{\eea}{\end{eqnarray}}
\newcommand{\ii}{\mbox{i}}
\newcommand{\pa}{\partial}
\newcommand{\vep}{\varepsilon}

\newcommand{\nn}{\nonumber}

\begin{document}

\maketitle

\begin{abstract}\small{
It is a well accepted principle that finite-energy  static
solutions in the classical relativistic gauge field theory over the 
$(2+1)$-dimensional Minkowski spacetime must be 
electrically neutral. We call such a statement the Julia--Zee theorem.
In this paper, we present a mathematical proof of this fundamental  
structural property.

\medskip


}
\end{abstract}

\section{Introduction}
\setcounter{equation}{0}

Consider the Maxwell equations
\be\label{Maxwell}
\pa_\nu F^{\mu\nu}= -J^\mu 
\ee
defined over a Minkowski spacetime of signature $(+-\cdots-)$, where 
\be 
F_{\mu\nu}=\pa_\mu A_\nu -\pa_{\nu}A_\mu 
\ee 
is the electromagnetic tensor induced from the gauge vector field $A_\mu$, $\mu=0$ designates the temporal index, $\mu=i, j,k$ denote the spatial indices,
and $J^\mu=(J^0,J^i)=(\rho,{\bf j})$ is the current density in which $\rho=J^0$ expresses the electric charge density. As spatial vector fields,
the electric field ${\bf E}=(E^i)$ and magnetic field ${\bf B}=(B^i)$ are given by
\be \label{F}
F^{0i}=-E^i,\quad F^{ij}=-\vep^{ijk} B^k.
\ee
In view of (\ref{F}), the $\mu=0$ component of (\ref{Maxwell}) relating electric field and charge density reads
\be\label{E}
\mbox{div}\, {\bf E}=\pa_i E^i=\rho,
\ee
which is commonly referred to as the Gauss law (constraint). In the static situation, we have
\be 
E^i=-F^{0i}=F_{0i}=-\pa_i A_0.
\ee
Thus, a trivial temporal component of the gauge field,
\be \label{temporalgauge}
A_0=0,
\ee
implies that electric field is absent, ${\bf E}={\bf 0}$. The condition (\ref{temporalgauge}) is also known as the temporal gauge 
condition, which makes the static solution electrically neutral.

In their now classic 1975 paper \cite{JZ}, Julia and Zee studied the Abelian Higgs gauge field theory model. Using a radially symmetric
field configuration ansatz  and assuming a sufficiently fast decay rate at spatial infinity, they were able to conclude that a 
finite-energy static solution
of the equations of motion over the $(2+1)$-dimensional Minkowski spacetime must satisfy the temporal gauge condition (\ref{temporalgauge}),
and thus, is necessarily electrically neutral. This result, referred here as the Julia--Zee theorem, leads to many interesting consequences.
For example, it makes it transparent that the static Abelian Higgs model is exactly the Ginzburg--Landau theory \cite{GL} which is purely
magnetic \cite{JT,NO}. Since the work of Julia and Zee \cite{JZ}, it has been accepted \cite{Dunne,HKP,JW,KK,PK,VS} that, in order to obtain both electrically and magnetically charged static vortices,
one needs to introduce into the Lagrangian action density the Chern--Simons topological terms \cite{CS1,CS2}, which is an essential construct in anyon physics \cite{W1,W2}. See also \cite{FM}.

On the other hand, it is well known that electrically and magnetically charged static solitons, called dyons by Schwinger \cite{Sch}
(see also the related work of Zwanziger \cite{Z1,Z2}), exist as
solutions to the Yang--Millis--Higgs equations over $(3+1)$-dimensional spacetime \cite{JZ,PS,SW}. Therefore, the Julia--Zee theorem is valid only in $(2+1)$ dimensions.

The importance of the Julia--Zee theorem motivated us to carry out this study. In  Section 2, we make a precise statement
of the Julia--Zee theorem  in the context of the original Abelian Higgs model and present a rigorous proof. In Section 3, we extend the
Julia--Zee theorem to the situation of a non-Abelian Yang--Mills--Higgs model. In Section 4, we prove a non-Abelian version of the theorem.
Fortunately our method works almost exactly as in the simpler Abelian Higgs model. In Section 5, we consider further extensions and applications
of our results.

\section{The Julia--Zee Theorem}
\setcounter{equation}{0}

Recall that, in normalized units, the classical Abelian Higgs theory over the $(2+1)$-dimensional spacetime is governed by the Lagrangian action density
\be\label{1.1}
{\cal L}=-\frac14 F_{\mu\nu}F^{\mu\nu}+\frac12 D_\mu\phi\overline{D^\mu\phi}-V(|\phi|^2)
\ee
where $D^\mu\phi=\pa^\mu\phi+\ii A^\mu \phi$ defines the gauge-covariant derivative, $\phi$ is
a complex scalar (Higgs) field, the spacetime indices $\mu,\nu$ run through $0,1,2$,  the spacetime metric takes the
form $\eta=(\eta_{\mu\nu})=\mbox{diag}(1,-1,-1)$, which is used to lower and raise indices, and $V\geq0$ is the potential density of
the Higgs field.
The associated equations of motion are
\bea
D_\mu D^\mu\phi&=&-2 V'(|\phi|^2)\phi,\label{2a}\\
\pa_\nu F^{\mu\nu}&=&-J^\mu,\label{2b}\\
J^\mu&=&\frac\ii2 (\overline{\phi}D^\mu\phi-\phi\overline{D^\mu\phi}).\label{2c}
\eea

In the static situation, the operator $\pa_0=0$ nullifies everything. Hence the electric charge density $\rho$ becomes
\be\label{3.1}
\rho=J^0=\frac\ii2 (\overline{\phi}D^0\phi-\phi\overline{D^0\phi})=-A_0|\phi|^2
\ee
and  a nontrivial temporal component of the gauge field, $A_0$, is necessary
for the presence of electric charge.

On the other hand, the $\mu=0$ component of the left-hand side of the Maxwell equation (\ref{2b}) is
\be\label{4}
\pa_\nu F^{0\nu}=\pa_i (F_{i0})=\pa_i^2 A_0=\Delta A_0.
\ee
Consequently, the static version of the equations of motion (\ref{2a})--(\ref{2c}) may be written as
 \bea
D_i^2\phi&=&2V'(|\phi|^2)\phi-A_0^2\phi,\label{5a}\\
\pa_j F_{ij}&=&\frac\ii2(\overline{\phi}D_i\phi-\phi\overline{D_i\phi}),\label{5b}\\
\Delta A_0&=&|\phi|^2 A_0,\label{5c}
\eea
in which (\ref{5c}) is the Gauss law.
Moreover, since the energy-momentum (stress)
tensor has the form
\be\label{7.1}
T_{\mu\nu}=-\eta^{\mu'\nu'}F_{\mu\mu'}F_{\nu\nu'}+\frac12(D_\mu\phi\overline{D_\nu\phi}
+\overline{D_\mu\phi}D_\nu\phi)-\eta_{\mu\nu}{\cal L},
\ee
the Hamiltonian density is given by
\be\label{8}
{\cal H}=T_{00}=\frac12|\pa_i A_0|^2+\frac12 A_0^2|\phi|^2+\frac14 F_{ij}^2
+\frac12|D_i\phi|^2+V(|\phi|^2),
\ee
so that the finite-energy condition reads 
\be\label{9}
\int_{\bfR^2}{\cal H}\,\dd x<\infty.
\ee
 
With the above formulation, the Julia--Zee theorem may be stated as follows.

\begin{theorem} (The Julia--Zee theorem) Suppose that $(A_0, A_i, \phi)$ is a finite-energy solution of the static Abelian Higgs equations
(\ref{5a})--(\ref{5c}) over $\bfR^2$. Then either $A_0=0$ everywhere if $\phi$ is not identically zero or $A_0\equiv$ constant and the solution is necessarily electrically neutral.
\end{theorem}

Our proof of the theorem is contained in the following slightly more general statement.

\begin{proposition} Let $A_0$ be a solution of $\Delta A_0 =|\phi|^2 A_0$  over $\bfR^2$. Suppose that
\be 
\int_{\bfR^2}|\nabla A_0|^2\, \dd x <\infty.
\ee
Then $A_0=$constant. Furthermore, if  $\phi$ is not identically zero, then $A_0 \equiv 0$.
\end{proposition}

\begin{proof}  Let $0 \leq \eta \leq 1 $ be of compact support and define for $M>0$ fixed the truncated function
\begin{equation}
A_0^M=\left\{ \begin{array}{rl}
M  & \mbox{if}~ A_0>M,\\
A_0 &\mbox{if}~ |A_0| \leq M, \\
-M & \mbox{if} ~A_0 <-M.
\end{array} \right.  \label{eq1} 
\end{equation}
 Then, multiplying (\ref{5c}) by $\eta A_0^M$ and integrating, we have
 \begin{equation}\label{eq2}
 \int_{\bfR^2}[\eta \nabla A_0\cdot \nabla A_0^M +A_0^M \nabla A_0 \cdot \nabla \eta +\eta  |\phi|^2 A_0^M A_0]
\,  \dd x=0.
\end{equation}

Using (\ref{eq1}) in (\ref{eq2}), we find
\bea
\label{eq3} 
&&\int_{\{|A_0|<M\}\cap~\mbox{supp}(\eta)} \eta |\phi|^2 A_0^2\, \dd x+  M^2\int_{\{|A_0|>M\}\cap~\mbox{supp}(\eta)}
\eta |\phi|^2\, \dd x \nn\\
&& + \int_{\{|A_0|<M\}\cap~\mbox{supp}(\eta)}\eta|\nabla A_0|^2\, \dd x \nn\\
 &\leq &
M\bigg(\int_{\bfR^2}|\nabla A_0|^2\, \dd x\bigg)^{\frac12}\bigg(\int_{\bfR^2}|\nabla \eta|^2\, \dd x\bigg)^{\frac12}.
\eea

For $R>0$, we now choose $\eta$ to be a logarithmic cutoff function given as
\begin{equation}
\label{eq4} \eta=\left\{ \begin{array}{ll}
1 & \mbox{if}~ |x|<R,\\
2-\frac{\log{|x|}}{\log R} &\mbox{if}~ R\leq |x|\leq R^2, \\
0  & \mbox{if}~ |x|>R^2.
\end{array} \right. 
\end{equation}

Then 
\be\label{int}
\int_{R^2}|\nabla \eta|^2\,\dd x=\frac{2\pi}{\log{R}}.
\ee
 Using (\ref{int}) in 
(\ref{eq3}) gives
\begin{eqnarray}
\label{eq5} 
&&\int_{\{|A_0|<M\}\cap B_R}|\phi|^2 A_0^2\, \dd x+\int_{\{|A_0|<M\}\cap B_R}|\nabla A_0|^2\, \dd x\nn\\
&\leq&
\int_{\{|A_0|<M\}\cap B_R}|\phi|^2 A_0^2\, \dd x+ M^2\int_{\{|A_0|>M\}\cap B_R}
|\phi|^2\, \dd x +\int_{\{|A_0|<M\}\cap B_R}|\nabla A_0|^2\,\dd x  \nn\\
&\leq& 
M\frac{\bigg(2\pi \int_{\bfR^2}|\nabla A_0|^2\, \dd x\bigg)^{\frac12}}{( \log R)^{\frac12}}.
\end{eqnarray}
The right hand side of (\ref{eq5}) tends to zero as $R$ tends to infinity. Letting $M$ tend to infinity
proves the proposition.
\end{proof}

\section{A Non-Abelian Julia--Zee Theorem}
\setcounter{equation}{0}
In this section, we consider the simplest non-Abelian Yang--Mills--Higgs theory for
which the gauge group is $SU(2)$ or $SO(3)$. For convenience, we work with the gauge group in the
adjoint representation so that the
gauge field and Higgs field are all real 3-vectors expressed as
\[
{\bf A}_\mu=(A_\mu^1,A^2_\mu,A_\mu^3)^T,\quad
\Phi=(\phi^1,\phi^2,\phi^3)^T.
\]
Then the Yang--Mills field curvature tensor and gauge-covariant derivative are given by
\[
{\bf F}_{\mu\nu}=\pa_\mu {\bf A}_\nu-\pa_\nu{\bf A}_\mu+\bA_\mu\times \bA_\nu,
\quad
D_\mu \Phi=\pa_\mu\Phi +{\bf A}_\nu\times \Phi,
\]
so that the Lagrangian density is written as
\bea\label{L}
{\cal L}&=&-\frac14 {\bf F}_{\mu\nu}\cdot {\bf F}^{\mu\nu}+\frac12 D_\mu\Phi\cdot D^\mu\Phi-V(\Phi)\nn\\
&=&-\frac14 {\bf F}_{ij}\cdot{\bf F}_{ij}+\frac12 {\bf F}_{0i}\cdot{\bf F}_{0i}-\frac12 D_i\Phi\cdot D_i\Phi
+\frac12 D_0\Phi\cdot D_0\Phi-V(\Phi).
\eea
As a consequence, the equations of motion, or the Yang--Mills--Higgs equations, are
\bea
D^\mu {\bf F}_{\mu i}+\Phi\times D_i\Phi&=&0,\label{YM}\\
D^\mu D_\mu \Phi+\frac{\delta V(\Phi)}{\delta\Phi}&=&0,\label{H}
\eea
coupled with the Gauss-law constraint
\be\label{1}
D^\mu {\bf F}_{\mu0}+\Phi\times D_0\Phi=0.
\ee
This is the equation that governs the ${\bf A}_0$ field and is of our main concern.
The actual form of (\ref{1}) is:
\be\label{2}
-D_i {\bf F}_{i0}+\Phi\times D_0\Phi=0,\quad i=1,2.
\ee
In the static case, $\pa_0=0$. So
\be\label{3}
D_0\Phi={\bf A}_0\times\Phi,\quad
{\bf F}_{i0}=\pa_i {\bf A}_0+{\bf A}_i\times {\bf A}_0.
\ee
Inserting (\ref{3}) into (\ref{2}), we get
\be\label{5}
\Delta{\bf A}_0+\pa_i ({\bf A}_i\times {\bf A}_0)+{\bf A}_i\times\pa_i{\bf A}_0
+{\bf A}_i\times ({\bf A}_i\times{\bf A}_0)-\Phi\times ({\bf A}_0\times\Phi)=0.
\ee

On the other hand, the Hamiltonian density of the theory is
\bea\label{6}
{\cal H}&=& {\bf F}_{0i}\cdot{\bf F}_{0i}+D_0\Phi\cdot D_0\Phi-{\cal L}\nn\\
&=&\frac14 {\bf F}_{ij}\cdot{\bf F}_{ij}+\frac12{\bf F}_{0i}\cdot{\bf F}_{0i}+\frac12 D_i\Phi\cdot D_i\Phi
+\frac12 D_0\Phi\cdot D_0\Phi+V(\Phi),
\eea
which is positive definite.
The terms containing ${\bf A}_0$ give us the ${\bf A}_0$ energy
\bea\label{7}
E({\bf A}_0)&=&\int_{\bfR^2}\bigg\{\frac12 {\bf F}_{0i}\cdot{\bf F}_{0i}+\frac12 D_0\Phi\cdot D_0\Phi\bigg\}\,\dd x\nn\\
&=&\int_{\bfR^2}\bigg\{\frac12|\pa_i{\bf A}_0+({\bf A}_i\times{\bf A}_0)|^2+\frac12|{\bf A}_0\times\Phi|^2\bigg\}\,\dd x.
\eea

It can be checked that the governing equation for ${\bf A}_0$, the equation (\ref{5}), is the Euler--Lagrange equation of
(\ref{7}).

\begin{theorem} (A non-Abelian extension of the Julia--Zee theorem) \label{theorem3.1} Let ${\bf A}_0$ be a solution of (\ref{1}) with finite energy $E({\bf A}_0)<\infty$.
Then $E({\bf A}_0)=0$.  In particular, ${\bf F}_{0i}\equiv{\bf 0}$, $D_0 \Phi\equiv{\bf0}$, and $|{\bf A}_0|\equiv$constant.
Moreover, if the nonnegative potential density $V$ is such that $V(\Phi)=U(|\Phi|^2)$,  $U(s)$ has its unique zero at some $s=\theta^2\geq0$, and
$(\Phi, \bA_i,\bA_0)$ is a finite-energy solution of the Yang--Mills--Higgs equations (\ref{YM})--(\ref{1}), then
$\bA_0\equiv0$.  Otherwise the solution triplet $(\Phi, \bA_i,\bA_0)$  must be trivial, i.e.
\be 
|\Phi|\equiv\theta,\quad \bF_{ij}\equiv{\bf0},\quad E(\bA_0)=0.
\ee
\end{theorem}

To see that the absence of the electric sector in the non-Abelian Yang--Mills--Higgs model is implied by the above theorem, recall that the 
't Hooft electromagnetic tensor \cite{t} (see also \cite{Ryder,Tch} for related discussion and extension) may be written as
\be 
F_{\mu\nu}=\frac1{|\Phi|}\Phi\cdot\bF_{\mu\nu}-\frac1{|\Phi|^3}\Phi\cdot(D_\mu\Phi\times D_\nu\Phi).
\ee
Hence $E^i=F_{0i}\equiv0$ if $E(\bA_0)=0$.

\section{Proof of the Non-Abelian  Julia--Zee Theorem}
\setcounter{equation}{0}

 Let $0 \leq \eta \leq 1 $ be of compact support and define for $M>0$ fixed the truncated vector field
\begin{equation}\label{neq1}
\bA_0^M=\left\{ \begin{array}{cl}
\bA_0 &\mbox{if}~ |\bA_0| \leq M,\\
\frac{M}{|\bA_0|}\bA_0 &\mbox{if}~|\bA_0|>M.
\end{array} \right.  
\end{equation}
 Then, using $\eta \bA_0^M$ as a test function, we obtain from (\ref{5}) the expression
 \bea\label{neq2}
 &&\int_{\bfR^2}\bigg\{\eta [\partial_i \bA_0^M\cdot  \partial_i \bA_0 -\bA_0^M \cdot \partial_i (\bA_i \times \bA_0)
  -\bA_0^M\cdot(\bA_i \times \partial_i \bA_0)\nn\\
 && -\bA_0^M\cdot (\bA_i \times (\bA_i \times \bA_0))
  +\bA_0^M\cdot ( {\Phi}\times  (\bA_0\times{\Phi}))]+\partial_i \eta\,\bA_0^M \cdot \partial_i \bA_0\bigg\}\,\dd x=0.
  \eea
  
 Using the definition of $\bA_0^M$ in (\ref{neq1}), we see that 
\bea
\bA_0^M\cdot (\partial_i \bA_i \times \bA_0)&=&0,\nn\\
 (\bA_i \times \bA_0^M)\cdot (\bA_i \times \bA_0)&=&\frac{|\bA_0|}{M}|\bA_i \times
 \bA_0^M|^2 \,\,~\mbox{in $\{|\bA_0|>M\}$},\nn\\
 (\bA_0^M\times{\Phi})\cdot (\bA_0\times {\Phi})&=&\frac{|\bA_0|}{M}|\bA_0^M \times {\Phi}|^2 \,\,~\mbox{in $\{|\bA_0|>M\}$},\nn\\
 \partial_i \bA_0^M \cdot \partial _i \bA_0&=&\frac{|\bA_0|}{M} (\partial_i \bA_0^M)^2 \,\,~\mbox{in $\{|\bA_0|>M\}$},\nn\\
 -2\bA_0^M\cdot(\bA_i\times \partial_i \bA_0)&=&2\frac{|\bA_0|}M \partial_i \bA_0^M \cdot (\bA_i\times \bA_0^M)
 \,\,~\mbox{in $\{|\bA_0|>M\}$}.\nn
\eea
 
 We then obtain from  (\ref{neq2}) that
 \bea\label{neq3}
 &&\int_{\{|\bA_0|\leq M\}}\eta\{|\partial_i \bA_0+(\bA_i\times \bA_0)|^2+|\bA_0\times {\Phi}|^2\}\,\dd x\nn\\
 &&+\int_{\{ |\bA_0|>M\}}\frac{|\bA_0|}M\eta \{|\partial_i \bA_0^M+\bA_i\times \bA_0^M|^2+|\bA_0^M \times
{\Phi}|^2\}\,\dd x\nn\\
 &=&-\int_{\bfR^2}\{\partial_i \eta\,\bA_0^M \cdot \partial_i \bA_0\}\, \dd x\nn\\
&=&
 -\int_{\bfR^2}\{\partial_i \eta\,\bA_0^M \cdot (\partial_i \bA_0+\bA_i\times \bA_0)\}\, \dd x.
 \eea

We again choose $\eta$ according to (\ref{eq4}).
 Using (\ref{int}), we have 
\bea\label{neq5} 
&&\int_{\{|\bA_0\leq M\}\cap B_R}\{|\partial_i \bA_0+(\bA_i\times \bA_0)|^2+|\bA_0\times {\Phi}|^2\}\,\dd x\nn\\
&\leq& M\bigg(\frac{2\pi}{ \log R}\bigg)^{\frac12}\bigg( \int_{\bfR^2}|\partial_i \bA_0+\bA_i \times \bA_0|^2\, \dd x\bigg)^{\frac12}\nn\\
&\leq&  M\bigg(\frac{4\pi }{ \log R}E(\bA_0)\bigg)^{\frac12}.
\eea
The right hand side of (\ref{neq5}) tends to zero as $R$ tends to infinity. Letting $M$ tend to infinity
proves $E(\bA_0)=0$.

To see that $|\bA_0|=$constant, we use the result ${\bf F}_{0i}={\bf0}$ to deduce that $\pa_i|\bA_0|^2=2 \pa_i\bA_0\cdot \bA_0=-2(\bA_i\times \bA_0)\cdot \bA_0=0$.

Suppose $\bA_0\neq{\bf0}$. Then $|\bA_0|=a>0$ for some constant $a$. Note that $E(\bA_0)=0$ also implies that $\Phi$ remains parallel to 
$\bA_0$ everywhere. So there is a scalar function $u$ such that $\Phi=u\bA_0$. Consequently, we have
\be\label{D} 
D_i\Phi=(\pa_i u)\bA_0 +u D_i \bA_0=(\pa_i u)\bA_0 +u \bF_{i0}=(\pa_i u)\bA_0.
\ee 

Now assume that the Higgs potential density takes the form $V(\Phi)=U(|\Phi|^2)$. Iterating (\ref{D}), we get $D_i D_i \Phi=(\Delta u)\bA_0$.
Hence, by (\ref{H}) and $D_0\Phi={\bf0}$, we arrive at
\be \label{equ}
\Delta u=U'(a^2 u^2 )u\quad\mbox{in }\bfR^2.
\ee
In view of the finite-energy condition and (\ref{D}), we have
\be \label{I}
I(u)=\int_{\bfR^2}\bigg\{\frac12|\nabla u|^2+\frac1{2a^2}U(a^2 u^2)\bigg\}\,\dd x<\infty.
\ee
It may easily be checked that, as a solution of (\ref{equ}), $u$ is a finite-energy critical point of the functional (\ref{I}). However, using a 
standard rescaling argument with
$x\mapsto x_\sigma=\sigma x$ and $u(x)\mapsto u_\sigma(x)=u(x_\sigma)$ so that $\dd I(u_\sigma)/\dd \sigma=0$ at $\sigma=1$, we find
\be 
\int_{\bfR^2} U(|\Phi|^2)\,\dd x=\int_{\bfR^2}U(a^2 u^2)\,\dd x=0,
\ee
which implies that $\Phi$ lies in the ground state, $|\Phi|\equiv\theta$. As a consequence, $u\equiv \pm \frac\theta a$ or
\be 
\Phi=\pm \frac \theta a \bA_0,
\ee
which immediately gives us $D_i\Phi=\pm\frac \theta a D_i \bA_0=\pm\frac\theta a\,\bF_{i0}={\bf0}$ over $\bfR^2$.

Therefore, the coupled equations (\ref{YM}) and (\ref{H}) are reduced to the pure static Yang--Mills equations
\be 
D_i \bF_{ij}={\bf 0}\quad\mbox{in }\bfR^2,
\ee
which is known to have only the trivial solution, $\bF_{ij}={\bf 0}$, over $\bfR^2$, as can easily be seen from a similar rescaling argument involving
$x\mapsto x_\sigma=\sigma x$ and $\bA_i(x)\mapsto (\bA_\sigma)_i (x)=\sigma \bA_i(x_\sigma)$, $i=1,2$, in the energy functional
\be 
\int_{\bfR^2}|\bF_{ij}|^2\,\dd x.
\ee

The proof of the stated non-Abelian extension of the Julia--Zee  theorem is complete.

\section{Extension and Application}
\setcounter{equation}{0}

As an extension, consider a general non-Abelian gauge group, say the unitary group $U(N)$, with  Lie algebra ${\cal U}(N)$ consisting of $N\times N$ anti-Hermitian matrices. Then 
\be 
\langle A,B\rangle =-\mbox{Tr}(AB),\quad A,B\in {\cal U}(N),
\ee
is the inner product over ${\cal U}(N)$ which allows one to regard the Lie commutator, $[\,,\,]$ on ${\cal U}(N)$, as an exterior product so that
\be \label{5.2}
\langle A,[A,B]\rangle =0,\quad \langle A, [B,C]\rangle= \langle C, [A,B]\rangle= \langle B, [C,A]\rangle,\quad A,B,C\in {\cal U}(N).  
\ee

The $U(N)$ Yang--Mills--Higgs theory with the Higgs field $\Phi$ represented adjointly has the Lagrangian action density
\be \label{5.3}
{\cal L}=-\frac14\langle F_{\mu\nu},F^{\mu\nu}\rangle +{\frac12} \langle D_\mu\Phi,D^\mu\Phi \rangle -V(\Phi).
\ee 

In view of the method in Section 4 and the property (\ref{5.2}), we may similarly show that a finite-energy static solution of the equations of motion of 
the Yang--Mills--Higgs theory in the $(2+1)$-dimensional Minkowski spacetime defined by (\ref{5.3}) has a trivial temporal component, $A_0$.

Furthermore, it is clear that our result applies to the models that contain several Higgs fields as well.

\medskip

As an application, consider the classical Abelian Chern--Simons--Higgs theory \cite{PK} defined by the Lagrangian action density
\begin{equation}\label{LCS}
{\cal L} = - \frac14 F_{\mu \nu} F^{\mu \nu} + \frac{\kappa}{ 4} \varepsilon^{\mu \nu \alpha} A_\mu F_{\nu \alpha} + \frac12 D_\mu \phi \overline{D^\mu \phi} - \frac{\lambda}{ 8} ( \left| \phi \right|^2 - 1 )^2,
\end{equation}
over the $(2+1)$-dimensional Minkowski spacetime, where $\kappa$ is the Chern--Simons coupling constant. The equations of motion governing static field configurations are
\bea
\label{cs1}
D_j^2 \phi &=& \frac\lambda2 (|\phi|^2-1)\phi-A_0^2\phi,\\
\label{cs2}
\partial_k F_{jk}-\kappa\,\varepsilon_{jk} \partial_k A_0 &=&\frac\ii2(\overline{\phi}D_j\phi-\phi\overline{D_j\phi}),\\
\Delta A_0 &=& \kappa F_{12} + |\phi|^2 A_0.\label{cs3}
\eea
Using the Julia--Zee theorem stated in Section 2 and the existence theorem obtained in \cite{Spirn}, we see that a finite-energy solution of the Chern--Simons--Higgs equations
(\ref{cs1})--(\ref{cs3}) exists which has a nontrivial temporal component $A_0$ of the gauge field, hence a nontrivial electric sector
is present in the theory, if and only if $\kappa\neq0$, which switches on the Chern--Simons topological term in the model. 

In view of Theorem \ref{theorem3.1}, similar applications may be made to non-Abelian Chern--Simons--Higgs vortex models \cite{KK,VS,VS1}.

\medskip
\medskip

{\bf Acknowledgments.}
The authors were supported in part by the NSF.

\end{document}